\documentclass[twocolumn,showpacs,prc]{revtex4}
\usepackage{dcolumn}
\usepackage{amsmath}
\providecommand{\LyX}{L\kern-.1667em\lower.25em\hbox{Y}\kern-.125emX\@}

\usepackage{graphics}
\makeatother
\begin{document}

\newcommand{\be}{\begin{eqnarray}}
\newcommand{\ee}{\end{eqnarray}}
\topmargin -1cm

\title{The continuum description with pseudo-state wave functions}

\author{A.~M. Moro}
\email{moro@us.es}

\author{M. Rodr\'{\i}guez-Gallardo}
\altaffiliation{
Present address: Centro de F\'{\i}sica Nuclear, Universidade de Lisboa, 
Av.\ Prof.\ Gama Pinto 2, 1649-003 Lisboa, Portugal}
\email{mrodri@cii.fc.ul.pt}

\affiliation{
Departamento de F\'{\i}sica At\'omica, Molecular y
Nuclear, Universidad de Sevilla,
Apartado~1065, E-41080 Sevilla, Spain}

\author{R. Crespo}
\email{Raquel.Crespo@tagus.ist.utl.pt}
\affiliation{Departamento de
F\'{\i}sica, Instituto Superior T\'ecnico, Taguspark,\\
Av.\ Prof.\ Cavaco  Silva,  
2780-990 Porto Salvo, Oeiras, Portugal}
\author{I.~J. Thompson}
\email{I.Thompson@surrey.ac.uk}
\affiliation{Departament of Physics,
University of Surrey, Guildford GU2 7XH, United Kingdom}

\date{\today}

\begin{abstract}
Benchmark calculations are performed aiming to test 
the use of two different pseudo-state bases on 
the Multiple Scattering expansion
of the total Transition amplitude scattering framework.  
Calculated  differential cross sections for 
$p$-$^{6}$He inelastic scattering at 717 MeV/u show a good agreement between the observables calculated in the two bases.
This result gives extra confidence on the pseudo-state representation of continuum states to describe inelastic/breakup scattering.

\end{abstract}

\maketitle


Inelastic scattering at intermediate energies can be a useful tool
to study multipole excitations of Borromean nuclei
(like $^{11}$Li and $^6$He). Due to their loosely bound nature, to properly 
understand and interpret such reactions, 
it is crucial to take into account the few-body degrees of
freedom. At  high energies, the Multiple Scattering expansion 
of the total Transition amplitude
(MST) is a convenient framework that has already been applied to
analyze such reactions for elastic \cite{Cres01,Cres06} 
as well as for inelastic \cite{Cres02,Cres06b} scattering. 
In the latter case, the method can take into account spin excitations
that occur when scattering from a spin target such as a proton.
In these calculations, it is formal and numerically advantageous to represent the continuum states in terms of a basis of square-integrable functions, 
also known as pseudo-states (PS). Unlike the true scattering states,
the PS vanish at large distances and hence the method will be only useful if 
the calculated observables are not sensitive to the asymptotic 
region. Moreover, calculations performed with different families of states, 
should converge to the same results, provide that enough states are included, 
and that the basis is complete within the radial region which is relevant 
for the process under study.

Guided by this motivation, in this Brief Report we present benchmark calculations
of proton inelastic scattering from $^6$He within the MST scattering framework
making use of two different PS bases to describe 
the $^6$He continuum. We aim to check to what extent the calculated breakup observables depend on the choice of the PS functions.



For a Borromean system, like $^{6}$He, the wave function 
for a total angular momentum
$J$ (with projection $M$) and energy $\epsilon$,  $\varphi^{JM}_{\epsilon}$,
can be expressed in terms 
of the Jacobi coordinates  $\vec{r}$ (the relative coordinate between the valence nucleons) 
and $\vec{R}$ (the relative coordinate from the center of mass of the neutron pair to the core).

It is also convenient to introduce a set of hyperspherical coordinates:
the hyperradius  $\rho$ and five hyperspherical polar angles
$\Omega_5 = \{\alpha, \theta_x, \phi_x, \theta_y, \phi_y\} $. The former
is defined as $\rho = \sqrt{x^2 + y^2}$
with scaled coordinates $\vec{x} = 2^{-1/2} \vec{r}$
and $\vec{y}=(2/\sqrt{3})\vec{R}$.
The angle $\alpha = \arctan(x/y)$ is the hyperangle
and $ \theta_x, \phi_x, \theta_y, \phi_y$ the angles associated
with the unit spatial vectors $\hat{x}$ and $\hat{y}$.

Within the PS method, the eigenstates  $\varphi^{JM}_{\epsilon}(\vec{r},\vec{R})$ 
are obtained by diagonalization of the Hamiltonian in a basis of normalizable states. These 
states are conveniently expanded in a basis of Hyperspherical Harmonics of the form
\be
\psi^{JM}_{n\beta}(\vec{r},\vec{R})
=  R_{n\beta }(\rho)
\Upsilon_{\beta }^{J M}(\Omega_5)~~,
\label{HHwf}
\ee 
where  ${\Upsilon}_{\beta}^{J M}(\Omega_5)$ is the generalized angle-spin basis 
\cite{Dan98}
\be
{\Upsilon}_{\beta}^{J M}(\Omega_5)  =
\left\{
{\cal Y}_{{\cal K}  \ell_x \ell_y L}(\Omega_5)
\otimes \left[ \chi_{s_2} \otimes \chi_{s_3} \right]_{S}
\right\}_{ J M } ~~
\ee
with $\chi_{s_i}$ the neutron spin functions and  
${\cal Y}_{{\cal K}  \ell_x \ell_y L}(\Omega_5)$ the hyperspherical harmonics,
\be
{\cal Y}_{{\cal K}  \ell_x \ell_y L M_L}(\Omega_5) =
\psi_{{\cal K}  \ell_x \ell_y }(\alpha)
 \left[ Y_{\ell_x}(\hat{x}) \otimes  Y_{\ell_y}(\hat{y}) \right]_{LM_L} ~.
\ee
The functions  $\psi_{{\cal K}  \ell_x \ell_y}(\alpha)$
have an explicit form in terms of Jacobi polynomials  of
the hyperangle $\alpha$ \cite{Dan98}.
The set of quantum numbers $\beta =\{{\cal K}\ell_x \ell_y L S \}$ defines a channel, with
$\ell_x$ and  $\ell_y$  the orbital angular momenta associated with the Jacobi coordinates $\vec{x}$ and $\vec{y}$, 
${\cal K}=\ell_x + \ell_y +2 \nu$ ($\nu$=0,1,2, $\ldots$)  the hyperangular momentum,
$\vec{L}=\vec{\ell}_x+\vec{\ell}_y$ the total orbital angular momentum and $S$ the spin of the particles related by the coordinate $\vec{x}$. In Eq.~(\ref{HHwf}), 
$R_{n\beta }(\rho)$ are the hyperradial functions and $n$ is an index that labels the basis states within a given channel $\beta$. 
These functions are orthogonalized such that
\be
\int_0^\infty d\rho \, \rho^5 R_{n\beta}(\rho)R_{n'\beta}(\rho) = \delta_{nn'}\ .
\ee



The aim of the present work is to compare two different choices
for the functions $R_{n\beta }(\rho)$ in the calculation of breakup observables within the MST scattering framework.
First, we consider the Gauss-Laguerre (GL) basis \cite{face}, whose hyperradial part, $R_n^{GL}(\rho)$, is given by
\be
R_n^{GL}(\rho) = {\rho_0}^{-3} [n!/(n+5)!]^{1/2} L_n^5(z) \exp(-z/2) \, ,
\ee
with $z=\rho/\rho_0$,  $L_n^5$ the generalized Laguerre polynomials,
and $\rho_0$ a parameter that sets the radial scale of the basis. 

The second choice is the Transformed Harmonic Oscillator (THO) basis, recently introduced in Ref.~\cite{gallardo05} for a three-body system. The THO method is based on the idea of transforming the bound ground state wave function of the system into the ground state wave function of the Harmonic Oscillator (HO), defining a Local Scale Transformation (LST). The ground state wave function can be written as a linear combination of the basis functions
(\ref{HHwf}), 
\be
\varphi^{J_0 M_0}_{0}(\vec{r},\vec{R})
=\sum_{\beta}  R_{\beta }^{\epsilon_0}(\rho)
\Upsilon_{\beta }^{J_0 M}(\Omega_5)~~,
\label{wfgs}
\ee
where we have introduced  the abbreviated notation $\varphi^{J_0 M_0}_{0} \equiv \varphi^{J_0 M_0}_{\epsilon_{0}}$. Then, the 
equation that defines the LST for each channel $\beta$ is
\be
\int_0^{\rho}d\rho'\rho'^5
| R^{{\epsilon_0}  }_{\beta}(\rho')|^2
=\int_0^{s}ds's'^5| R^{HO}_{0{\cal K}}(s')|^2,
\ee
where $R^{HO}_{0{\cal K}}(s)$ is the hyperradial part of the  HO ground state for the hyperangular momentum ${\cal K}$.
Then, the THO basis is constructed for each channel applying the LST, $s_{\beta}(\rho)$, to the HO basis
\be
 R^{THO}_{n\beta}(\rho)
=  R^{{\epsilon_0}}_{\beta}(\rho)
     L_n^{{\cal K}+2} \left( s^2_{\beta}(\rho) \right) ,
\ee
where $L_n^{{\cal K}+2}$ are generalized Laguerre polynomials of degree $n$. 
For channels not included in the ground state, information from one of the known (ground state) channels with the closest quantum labels to the channel of interest is used to construct the LST, as explained in Ref.~\cite{gallardo05}.

Neither the GL nor the THO functions are eigenstates of the Hamiltonian, but they provide a complete and orthonormal set in which the Hamiltonian can be diagonalized. For this purpose, the basis is truncated by setting a maximum value of the index $n$ ($n=0,\ldots,N$) as well as a maximum hyperangular momentum 
${\cal K}_\mathrm{max}$.
Upon diagonalization in the truncated basis, the eigenstates are obtained as
\be 
\varphi_{\epsilon_i}^{JM}(\vec{r},\vec{R})=
\sum_{n\beta}C_{n\beta}^{J \epsilon_i}\psi_{n\beta}^{JM}(\vec{r},\vec{R}).
\label{eigen}
\ee
where $\{\epsilon_i \}$ are their associated eigenvalues.


From the derivation above, it becomes apparent that the GL basis is obtained in a more straightforward way than the THO basis. However, the latter has some appealing properties that could make it more  suitable in some situations. In particular, the THO basis has the advantage of being constructed from the ground state wave function of the system. Thus, when we diagonalize the Hamiltonian in a finite THO basis,  the ground state is recovered for any size of the basis. By contrast, in the GL representation a large basis may be required to obtain a good description of the ground state.
Also, note that the hyperradial part of the GL basis is the same for all the channels $\beta$ while in the THO basis a different hyperradial part is calculated for each channel, with the correct  behavior at the origin ($\rho^{{\cal K}}$).


%

For a meaningful comparison between the two bases, we 
use the same three-body Hamiltonian to generate the GL and  THO eigenstates for $^6$He. In particular, we use the
n-n potential of Gogny, Pires  and Tourreil \cite{Gogny70}
with spin-orbit and tensor components
and we take the $n$-$^4$He  potential from \cite{Bang79}. Besides the pairwise
interactions, an effective three-body potential is included, with 
matrix elements of  the form  \cite{Dan98}
\be
V^{\rm 3B}_{ \beta' \beta }(\rho) =
\frac{ \delta_{\beta' \beta}V^{\rm 3B}_J }
{1+(\rho/5)^3} ~~.
\ee
The $J$=0 strength of this effective potential is tuned to reproduce 
the experimental three-body separation energy  and the $J>0$ strength is 
adjusted to obtain the $2^+_1$ resonance at the experimental energy.
 



We now consider the scattering process of  $^{6}$He, originally in its ground 
state, $| \varphi_{0}^{J_0 M_0}   \rangle$,
to a final continuum state $ |\varphi_{\epsilon}^{J M} \rangle$, at excitation energy $\epsilon$ and with total angular momentum $J$ (projection $M$), 
by means of its
interaction with a proton, with initial (final)
linear momentum $\vec{k}_1$ ($\vec{k}\,'_1$) in the nucleon-nucleus 
center-of-mass frame and spin $S_1=1/2$ 
with projection $\sigma$ (${\sigma}'$).

The double differential cross section for this process 
can be formally expressed as
\be
\label{dsdwde}
\frac{d^2 \sigma_{JJ_0}}{d \Omega d\epsilon}
&=& \frac{1}{( \widehat{S_1})^2}  \frac{1}{( \widehat{J_0})^2}
\left[\frac{\hbar^2}{4 \pi^2 \mu_{NA}}\right]^2
\nonumber \\
 &\times&  \sum_{\sigma {\sigma}'} \sum_{M M_0} 
|\langle \vec{k}'_{1} \chi_{S_1}^{{\sigma}'};
\varphi_{\epsilon}^{J M}
|{\cal T}|
\vec{k}_{1}\chi_{S_1}^{\sigma} ;\varphi_{0}^{J_0 M_0}  \rangle|^2
\nonumber \\
\ee
where ${\cal T}$ denotes the transition amplitude operator \cite{Joa87}. This operator can
be expressed as a multiple expansion series in the transition amplitudes 
$\hat{t}_{\cal I}$ for proton scattering from each projectile 
sub-system ${\cal I}$  \cite{Wat57}. At high energies 
and for small momentum transfers,
this expansion is expected to converge quickly. If only the leading term of the
series is retained, 
the single scattering approximation (SSA) is obtained \cite{Cres99,Cres02}:  
\begin{equation}
{\cal T}^{SSA} = \sum_{{\cal I}=2}^{4}\hat{t}_{1{\cal I}}  
\label{TMSexp}
\end{equation}
with ${\cal I}$=2,3 for the halo neutrons, and ${\cal I}$=4 for the core. The
proton - ${\cal I}$ subsystem transition amplitude satisfies
the Lippmann-Schwinger equation
\begin{equation}
\hat{t}_{1 {\cal I}} = v_{1 {\cal I}} + v_{1 {\cal I}}
 G_0 \hat{t}_{1 {\cal I}} ~~,
\end{equation}
with $v_{1 {\cal I}}$ the interaction between the nucleon and
${\cal I}$ subsystem.
Within the impulse approximation, the propagator $G_0 =\left( E^+ - K \right)^{-1}$
contains the
kinetic energy operators of the proton and all the projectile subsystems.
Here $E$ is the kinetic energy, $E = \frac{\hbar^2 k_1^2}{2 \mu_{NA}}$ in the
overall center of mass frame, and  $\mu_{NA}$ is the proton-projectile
reduced mass.

Within the PS method, the scattering states  $ \varphi_{\epsilon}^{J M}$ in Eq.~(\ref{dsdwde}) 
are approximated by the pseudo-states  $ \varphi_{\epsilon_i}^{J M}$. 
Hence, the double differential cross section (\ref{dsdwde}) becomes a single differential
cross section for each pseudo-state,
{\setlength\arraycolsep{1pt}
\be
\frac{d \sigma_{JJ_0}^{i}}{d \Omega }
&=& \frac{1}{( \widehat{S_1})^2}  \frac{1}{( \widehat{J_0})^2}
\left[\frac{\hbar^2}{4 \pi^2 \mu_{NA}}\right]^2 
 \nonumber \\
 &\times& 
\sum_{\sigma {\sigma}'} \sum_{M M_0}
|\langle \vec{k}'_{1} \chi_{S_1}^{{\sigma}'};
\varphi_{\epsilon_i}^{J M}
|\sum_{{\cal I}=2}^{4} \hat{t}_{1{\cal I}}|
\vec{k}_{1}\chi_{S_1}^{\sigma} ;\varphi_{0}^{J_0 M_0}  \rangle|^2
\nonumber \\
\ee}
\noindent
where we have replaced  the ${\cal T}$ matrix operator by its single scattering 
approximation.  Making use of the 
impulse approximation  \cite{Cres06}, the matrix elements for the scattering for each 
constituent can be further simplified, leading to the following factorized form 
for the scattering from one valence nucleon (${\cal I}$=2):
%
\be
\label{eq:sscatv}
& &\langle  \vec{k}'_{1} \chi_{S_1}^{{\sigma}'};\varphi_{{\epsilon}_i}^{J M}
|\hat{t}_{12}|\vec{k}_{1}\chi_{S_1}^{\sigma} ;\varphi_{0}^{J_0 M_0}  \rangle
  = 
\nonumber \\
 & & \sum_{b \beta}
 {t}_{[b \beta {\cal S}_p  {\cal S}'_p ]} (\omega_{12},{\Delta})
\times  \rho_{[b \beta ; {\cal S}_T {\cal S}'_T {\epsilon}_i]}
\left( \frac{m_3}{M_{23}}\vec{\Delta },\frac{m_4}{M_{234}}\vec{\Delta }\right)
\nonumber \\
\ee
with $ M_{23}=m_2 + m_3, M_{234}= m_2 + m_3 + m_4$ and where we have introduced the momentum transfer  $\vec{\Delta} =\vec{k}\,'_1 - \vec{k}_1$ and the energy parameter $\omega_{12}$ \cite{Cres06} 
and where ${\cal S}_p = \{ S_1 \sigma \} $ (${\cal S}'_p = \{ S_1 \sigma' \}$) 
are the incoming (outgoing) spin of the nucleon and its projection
and,
${\cal S}_T = \{ J_0 M_0 \}$ (${\cal S}'_T =  \{J M \}$)
the initial (final) total spin of the halo valence pair.
The amplitude
$ {t}_{[b \beta {\cal S}_p  {\cal S}'_p ]}$
is given in terms of the tensor components of the nucleon-nucleon
transition amplitude \cite{Cres06, Cres02b}. The transition density form 
factors,
$ \rho_{[b \beta ; {\cal S}_T {\cal S}'_T {\epsilon}_i]}$,
depend exclusively on the structure of the composite system. Its explicit 
expression as a function of the 
hyperradial parts of the  wave functions of the initial and final states, 
can be found in \cite{Cres06}.

The scattering from the core, assumed here as spinless, can 
equivalently be written as:
\be
\label{eq:sscatc}
& &\langle  \vec{k}'_{1} \chi_{S_1}^{{\sigma}'};\varphi_{{\epsilon}_i}^{J M}
|\hat{t}_{14}|\vec{k}_{1}\chi_{S_1}^{\sigma} ;\varphi_{0}^{J_0 M_0}  \rangle
  =  \nonumber \\
& & {t}_{[0 0 {\cal S}_p  {\cal S}'_p ]} (\omega_{14},{\Delta})
\times  \rho_{[0 0 ; {\cal S}_T {\cal S}'_T {\epsilon}_i      ]}
\left(0,\frac{M_{23}}{M_{234}}\vec{\Delta }\right)
\ee
where, as before, $\omega_{14}$ is the appropriate energy parameter \cite{Cres06}.
The angular differential cross section for $^{6}$He inelastic scattering (breakup) 
is then obtained by  summing all excited states contributions,
\be
\label{dsdw}
\frac{d \sigma^{inel}_{  JJ_0}}{d \Omega } &=&
\sum_{{\epsilon}_i}^{{\epsilon}_i^{\rm max}}
   \frac{d \sigma^{i}_{  JJ_0}}{d \Omega } \, . 
\ee 


For the evaluation of  Eqs.~(\ref{eq:sscatv}) and (\ref{eq:sscatc})
one needs the (free) transition amplitudes for
proton scattering from the valence nucleons and the core.
For the former, we used the NN Paris interaction. The transition 
amplitude for the 
$\alpha$ core was generated from  a phenomenological optical potential, of Woods Saxon form,
with parameters obtained by fitting existing data for the elastic scattering of
$p+^4$He  at  $E_{\rm p}=700$ and 800 MeV, as detailed in Ref.~\cite{Cres06}.

\begin{figure}
{\par\centering \resizebox*{0.45\textwidth}{!}
{\includegraphics{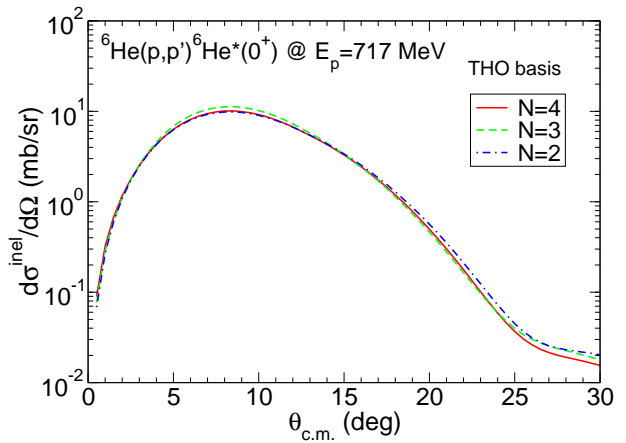}} \par}
\caption{\label{fig:converg}
 (Color online) Angular differential cross section for the breakup of $^6$He on protons 
at 717~MeV
per nucleon, leading to  $J^{\pi}=0^+$ continuum states of the $^6$He nucleus. The three 
lines represent the calculation with the THO basis, for several values of 
$N$ (indicated by the labels).  Each
curve includes the contribution from eigeinstates up to 10 MeV in excitation energy, according to 
Eq.~(\ref{dsdw}).
}
\end{figure}

We first study the convergence of the breakup observables with respect to the basis 
size. For this purpose, we consider the THO basis, truncated at different values of $N$. 
The maximum hyperangular momentum was set to
 ${\cal K}_\mathrm{max}=20$. This yields the three-body force parameters
$V_0^{\rm 3B}=-2.4$~MeV, for $J=0$, and $V_J^{\rm 3B}=-0.85$~MeV, for $J>0$.

In Fig.~\ref{fig:converg} we show the angular distribution of the  
calculated inelastic differential cross sections. 
For simplicity, 
only the $J^{\pi}=0^+$ continuum is included, and the Coulomb interaction 
between the proton and the $\alpha$ core is ignored.   The three lines 
represent the SSA calculation 
for different values of the basis size, according to the choice of the 
parameter $N$. The three cases are in almost perfect agreement, indicating that 
in this reaction the convergence with the basis size is very fast.

  
\begin{figure}
{\par\centering \resizebox*{0.45\textwidth}{!}
{\includegraphics{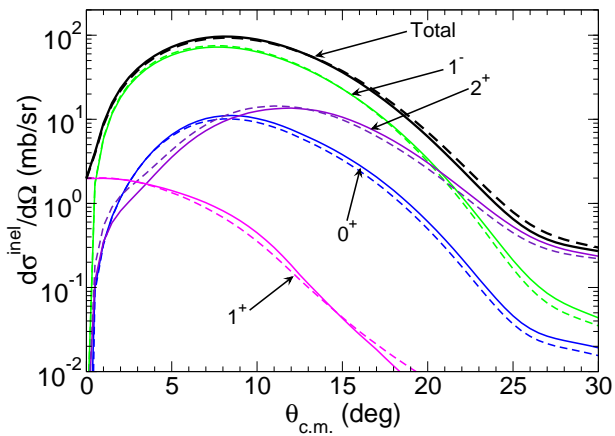}} \par}
\caption{\label{Fig:xsinelth-700}
 (Color online) Calculated contributions for breakup differential cross section 
leading to final states with $J^{\pi}=0^+, 1^{-}, 1^{+}$ and $2^+$, using the SSA approximation. 
Solid and dashed lines refer to the GL and THO basis, respectively. 
}
\end{figure}

Next, we study the dependence of the breakup observables on the choice of the basis, 
by comparing the calculations in the GL and THO representations.   
As before, the maximum hyperangular momentum was set to ${\cal K}_\mathrm{max}=20$, and 
only eigenstates below 10 MeV are considered.  The index 
$n$ was truncated to $N=20$ and $N=4$ for the GL and THO bases, respectively. With this model space, 
the number of pseudo-states in the GL (THO) basis is: 31 (30) for $0^+$,
63 (86) for  $1^-$, 53 (49) for $1^+$ and  79 (81) for $2^+$. For the GL basis, the range parameter 
was set to $\rho_0=0.25$~fm, which provides a basis that extends up to about 20~fm in the hyperradius.
With these parameters, the ground state obtained after diagonalization of the 
Hamiltonian appears at -0.9781 MeV and -0.9549 MeV, for the GL and THO bases, respectively.

In Fig.~\ref{Fig:xsinelth-700} we compare the inelastic angular distributions 
calculated in the GL and THO bases. The separate contributions for
$J^\pi$=$0^+$, $1^-$, $1^+$ and  $2^+$ final states are also shown. As before, the 
Coulomb interaction is neglected.  The thick 
lines are the incoherent sum of all these $J^\pi$ contributions. Solid and 
dashed lines correspond, respectively,
to the calculations with the GL and THO bases. For each curve, the contribution of eigenstates up 
to $\epsilon_{\max}=10$~MeV are added incoherently, according to Eq.~(\ref{dsdw}). We consider only 
the forward angles $\theta_\mathrm{c.m.}<30^{\circ}$ since the SSA is not expected to work well at large 
momentum transfers \cite{Cres06}. We see that, at these angles, the dominant contribution to 
the breakup cross section comes from the $1^-$ states, while  for  
$\theta_\mathrm{c.m.}>25^{\circ}$, the 
$2^+$ excitation becomes dominant. Finally,  the population of the unnatural $1^+$ states is 
almost negligible at all angles. We notice that this excitation mode requires spin-flip transitions which, according to these calculations, are very small in this reaction. 

For all these contributions, the GL and THO bases provide very similar results, 
suggesting that the calculated observables do not depend on the choice of the 
continuum representation, provided that enough states are included. 

In summary, in this  Brief Report  
we have calculated  proton inelastic scattering from $^6$He at 717 MeV/u,
using as scattering framework the single-scattering approximation and two different pseudo-state representations of
the  $^6$He continuum: the GL and the THO. Provided that 
enough states are included, both
bases predict essentially the same inelastic differential
cross section. Furthermore, the studied observables  converge very quickly with 
the size of the basis. These results support the reliability of the pseudo-state method as a 
useful and convenient tool to treat scattering problems dealing with continuum states. 
This  analysis could be extended to other PS bases and reactions.
 Furthermore, it could be
applied to other scattering frameworks, for which the PS method has been also 
implemented, such as the continuum discretized coupled channels (CDCC) 
method \cite{Mor06,Mat04a}.


{\bf Acknowledgements:}
This work was supported by the
Funda\c c\~ao para a Ci\^encia e Tecnologia
(Portugal) through grant No.\ POCTI/1999/FIS/36282, 
by the Acci\'on Integrada  HP2003-0121
and in the U.K.\ by EPSRC grant GR/M/82141. A.M.M. 
acknowledges a research grant  by 
the Junta de Andaluc\'{\i}a. We are grateful to J. G\'omez-Camacho 
and J.M. Arias for useful discussions.

\bibliographystyle{apsrev}

\end{document}